\documentclass[12pt]{amsart}

\usepackage{amssymb}
\usepackage{epsfig}
\usepackage{comment}
\usepackage{amsmath}
\usepackage{caption}
\usepackage[section]{placeins}
\usepackage{graphicx}
\usepackage{units}
\usepackage{color}
\usepackage{subfig}
\usepackage{threeparttable}
\usepackage[margin=1.0in]{geometry}
\graphicspath{ {Figures/} }

\theoremstyle{definition}

\numberwithin{equation}{section}
\theoremstyle{remark}

\begin{document}

\title[]{Uncertainty Quantification of Material Properties in Ballistic Impact of Magnesium Alloys}

\author[Xingsheng~Sun]
{Xingsheng~Sun${}^\dagger$}

\address
{
    Department of Mechanical and Aerospace Engineering,
    University of Kentucky,
    Lexington, KY 40506, USA.
    ${}^\dagger$ Corresponding author. E-mail address: xingsheng.sun@uky.edu.
}

\begin{abstract}
The design and development of cutting-edge light materials for extreme conditions including high-speed impact remains a continuing and significant challenge in spite of steady advances. Magnesium (Mg) and its alloys have gained much attention, due to their high strength-to-weight ratio and potential of further improvements in material properties such as strength and ductility. In this paper, we adopt a recently developed computational framework to quantify the effects of material uncertainties on the ballistic performance of Mg alloys. The framework is able to determine the largest deviation in the performance measure resulting from a finite variation in the corresponding material properties. It can also provide rigorous upper bounds on the probability of failure using known information about uncertainties and the system, and then conservative safety design and certification can be achieved. We specifically focus on AZ31B Mg alloys, and assume that the material is well-characterized by the Johnson-Cook constitutive and failure models, but the model parameters are uncertain. We determine the ordering of uncertainty contributions for model parameters and the corresponding behavior regimes where those parameters play a crucial role. Finally, we show that how this ordering provides insight on the improvement of ballistic performance and the development of new material models for Mg alloys.
\end{abstract}

\maketitle


\section{Introduction}
\label{sec:intro}

The assessment and design of armor plates for protecting humans and vehicles against high-speed impact has long been of interest in military and aerospace applications. Among a large number of protection materials, magnesium (Mg) and its alloys have gained much attention, due to their high strength-to-weight ratio and potential of further improvements in material properties such as strength and ductility~\cite{lloyd2021overview, wei2021insights, prameela2022strengthening}. The density of Mg is approximately $35\%$ lower than that of aluminum and approximately $77\%$ lower than that of steel. Therefore, Mg alloys are the lightest metallic material that have high potential for weight reduction, thereby decreasing the amount of fuel used in military and aerospace applications~\cite{dhari2001laser, mathaudhu2016magnesium}. However, compared with other conventional materials, such as aluminum and steel, fewer studies have been performed on the relationship between material properties of Mg alloys and its ballistic response, particularly under extreme conditions, i.e., high strain-rates and temperatures~\cite{eswar2014dynamic, sun2018proliferation, ponga2016dynamic}.

The ballistic impact of materials and structures is characterized by a very complex mechanical-thermal coupled process which mainly depends on the material properties and geometrical parameters of the target and the projectile, e.g., strength, toughness, shape and size~\cite{mohagheghian2015impact, jannotti2021role, liu2018ballistic}. These extreme complexities, which involve high non-linearity, singularity and dependence on a large number of parameters, render it impossible to derive a closed-form analytical solution. As a result, numerical methods, e.g., finite element methods, have attracted more attention and been employed extensively in the modeling of ballistic impact problems~\cite{xue2010penetration, li2012verification, li2013large}. In the numerical modeling of ballistics, the responses of the materials to mechanical and thermal loading conditions are always described by constitutive and failure models, which supplies a stress-strain relation at multiple temperatures and strain rates in order to formulate governing equations along with the kinematic and conservation laws. These constitutive and failure models are usually {\it empirical}, representing the primary link between experimental inputs and predicted outputs and hence constituting the strongest source of physical fidelity in a given calculation. The data used to calibrate material models can be achieved either traditionally by laboratory experiments for simple specimens such as split Hopkinson bar~\cite{ulacia2011tensile, ghosh2017plastic}, or by sub-grid scale simulations such as crystal plasticity~\cite{zhang2012phenomenological, chang2015variational}. Regardless of the source, analysts are most commonly restricted to a limited set of model forms, either by their simulation tool of choice or by the significant effort required to formulate, implement and characterize a new model. In practice, this restriction limits how well a fixed set of material parameters can fully represent a broad range of complex constitutive and failure behaviors. Therefore, the model parameters, characterizing multiple types of material properties, must be allowed to vary over a certain range, due to many factors including the potential complexity of response, any stochastic response characteristics, and the paucity of experimental data. These uncertainties render deterministic analysis of limited value. Instead, it becomes necessary to estimate the likely spread of performance metrics and relevant design requirement in order to provide an adequate design margin and meet specifications with sufficient confidence in the modeling of ballistic problems.

Within computational science, uncertainty quantification (UQ) is a family of powerful solution strategies that aim to characterize the variability of a given analysis and the spread in the predicted performance of a system~\cite{kovachki2022multiscale, gao2020bi, jiang2013novel, liu2018forward, liu2015dynamic}. The work presented in this paper focuses on systems in which the main source of uncertainty is an imperfect knowledge of material properties, as described by parameterized constitutive and failure models. The specific approach by which we quantify uncertainties is through the largest deviation in the performance measure resulting from a finite variation in the corresponding material properties, due to McDiarmid~\cite{mcdiarmid1989method} belong to a general class known as concentration-of-measure (CoM) inequalities~\cite{ledoux:01}. We also determine upper bounds of the probability that the system fails to perform within the design margin. Such bounds are rigorous, i.e., they are sure to be conservative and result in safe designs. These bounds are also optimal if we leverage all the known information about uncertainties and system using the optimal uncertainty quantification (OUQ) strategy~\cite{Owhadi:2013}. This characterization of systems using CoM and OUQ requires only knowledge of limiting ranges of the input variables, and not their full probability distribution as is the case of Bayesian methods. The upper bound performance characteristics are computed by exercising an existing deterministic code in order to sample the mean response of the system and to calculate the largest deviation in output in order to identify worst-case combinations of parameters. The CoM and OUQ strategies have been used in various applications including design of a thermal-hydraulic reactor~\cite{stenger2020optimal}, design of a fractal electrical circuit~\cite{topcu2011rigorous}, ballistic impact of aumnium alloys~\cite{kamga2014optimal, kidane2012rigorous, adams2012rigorous} and sub-ballistic impact of Mg alloys~\cite{sun2020rigorous, liu2021hierarchical}.

The AZ31B type of Mg alloys is chosen in this study since AZ31B is widely used in aerospace and automotive applications~\cite{jones2007ballistic, liu2008modelling, alderliesten2008applicability, sun2021concurrent}. The goal of this work is to investigate how the uncertainties in the constitutive and failure properties of AZ31B affect its ballistic performance when the material is subject to high-speed impact, and then to provide insight on how to improve such performance and develop new material models for AZ31B. For simplicity, we specifically hypothesize that the material behavior of AZ31B is known to be well-described by the Johnson-Cook constitutive and failure models~\cite{johnson1983constitutive, johnson1985fracture}, but the corresponding material constants are only imperfectly characterized and within speficic ranges. Then, both the resulting uncertainty in the performance and the mean performance measure are computed using the DAKOTA Version $6.7$ software package~\cite{adams2014dakota} of the Sandia National Laboratories. For all simulation parameters not considered as random variables, the values are considered as specified simulation conditions for evaluation. This strategy of fixing the boundary and initial conditions for assessment is entirely analogous to design testing for impact resistance~\cite{standard20080101}, wherein performance is evaluated relative to a targeted set of pre-characterized impact conditions. Simulations of such conditions are carried out over a range of impact velocities using the commercial finite-element package LS-DYNA~\cite{hallquist2007ls} on a single converged mesh. The ballistic tests conform to the form of the Recht-Ipson model, and the material failure mechanisms for AZ31B are noteworthy including spalling, plugging, discing and fragmentation. 

The remainder of the paper is structured as follows. In Section~\ref{sec:method} we start by reviewing the CoM inequality, the  OUQ approach and the corresponding probability bounds used for purposes of UQ. In Section~\ref{sec:sim}, we proceed to investigate the effects of material uncertainties on the ballistic performance of AZ31B Mg alloys subject to high-speed impact. We conclude with a summary and short discussion in Section~\ref{sec:concl}.

\section{Methodology}
\label{sec:method}

For the sake of completeness and convenience, in this section we briefly summarize the CoM and OUQ theories for rigorous UQ and conservative safe design. Additional details can be found in Refs.~\cite{lucas2008rigorous, Owhadi:2013, sun2020rigorous}.

\subsection{Concentration-of-measure (CoM) inequality}

We consider a system characterized by $N$ real-valued random variables $X \equiv (X_1,\dots,X_N) \in \mathbb{R}^N$ and a single real-valued performance measure $Y \in \mathbb{R}$.  The values of the input random variables lie within intervals $I \equiv (I_1,\dots,I_N)$, i.e., $x_1 \in I_1$, $\dots$ $x_N \in I_N$. We begin by supposing that the system performance can be described by a deterministic response function $F: \mathbb{R}^N \to \mathbb{R}$ through either experiments or an exact model. We further assume that the system fails if $Y \geq Y_c$ where  $Y_c$ is a threshold value for the safe operation of the system, and that the expected system performance $\mathbb{E} [{Y}]$ is known exactly. Then, a direct application of McDiarmid's concentration-of-measure (CoM) inequality~\cite{mcdiarmid1989method} provides an upper bound on the probability-of-failure of the system~\cite{sun2020rigorous}
\begin{equation} \label{eq:bound}
    \mathbb{P}[Y \geq Y_c]
    \leq
    \exp \bigg( -2\frac{\big(Y_c-\mathbb{E} [{Y}]\big)_+^2}{D^2} \bigg)
    \equiv
    {\rm P}_{\rm UB} ,
\end{equation}
where 
\begin{equation} \label{eq:diam}
    D = \Bigg( \sum_{i=1}^N D_i^2 \Bigg)^{1/2} 
\end{equation}
is the system diameter and $x_+:=\max(x,0)$.  In eq.~(\ref{eq:diam}), $D_i$ denotes the sub-diameter corresponding to the input variable $X_i$, which is calculated through the optimization problem
\begin{equation} \label{eq:individualdiam}
    D_i
    =
        \sup_{\hat{x}_i \in \hat{I}_i, \ x_i, x'_i\in {I}_i}
        \big| F(\hat{x}_i, x_i) - F(\hat{x}_i, x_i') \big|,
\end{equation}
where
\begin{subequations}
\begin{align}
 \hat{x}_i &= (x_1,\ldots,x_{i-1},x_{i+1},\ldots,x_N), \\
 \hat{I}_i &= {I}_1\times\cdots\times {I}_{i-1}\times {I}_{i+1} \times \cdots \times {I}_N, \\
 (\hat{x}_i, x_i) &= (x_1,\ldots,x_{i-1}, x_i,x_{i+1},\ldots,x_N), \\
 (\hat{x}_i, x_i') &= (x_1,\ldots,x_{i-1}, x_i',x_{i+1},\ldots,x_N).
 \end{align}
 \end{subequations}
 
The preceding methodology can be extended to the case in which the exact mean performance $\mathbb{E} [{Y}]$ is not available and the mean performance must be estimated instead. To this end, suppose that we conduct $n$ evaluations of the model $F({X})$ based on unbiased sampling of the random input variables, resulting in predicted performance measures $y^1$, $y^2$, ... , $y^n$. Then we can define an empirical mean performance as
\begin{equation}\label{f5swLP}
    \langle {Y} \rangle = \frac{1}{n} \sum _{k=1}^n y^k.
\end{equation}
Lucas et al.~\cite{lucas2008rigorous} showed that the probability of failure $\mathbb{P}[{Y} \geq Y_c ]$ can be determined to within confidence intervals by considering the randomness of the estimated mean $\langle {Y} \rangle$, with the result
\begin{equation} \label{eq:pofinterval}
    \mathbb{P} \Bigg[
        \mathbb{P} [Y \geq Y_c]
        \geq
        \exp\bigg( -2\frac{ \big(Y_c - \langle {Y} \rangle - \alpha \big)_+^2}{D^2}
    \bigg) \Bigg]
    \leq
    \epsilon',
\end{equation}
where $\epsilon'$ denotes a pre-specified tolerance for the mean estimation and $D$ is the same as in eq.~(\ref{eq:diam}). In eq.~(\ref{eq:pofinterval}), $\alpha$ characterizes the effect of estimating mean performance, which has a form of 
\begin{equation} \label{eq:marghit}
    \alpha= D \sqrt{\frac{-\ln \epsilon'}{2n}}.
\end{equation}
Another equivalent expression of eq.~(\ref{eq:pofinterval}) is that, with a probability that is greater than $1-\epsilon'$, we have
\begin{equation}\label{yuS4aF}
    \mathbb{P} [Y \geq Y_c]
    \leq
    \exp\Bigg( -2\frac{ \big(Y_c - \langle {Y} \rangle - \alpha \big)_+^2}{D^2} \Bigg) 
    \equiv
    {\rm P}_{\rm UB},
\end{equation}
which also supplies an upper bound on the probability of failure for the scenario of estimating mean performance.

With these identifications, a conservative and rigorous design criterion can be achieved through requiring that this upper bound on the probability of failure less than a tolerance, with the result
\begin{equation} \label{eq:certcrit}
    \text{CF} 
    \equiv 
    \frac{M}{U} 
	\equiv
	\frac{ \big(Y_c - \langle {Y} \rangle - \alpha \big)_+}{D}
    \geq 
    \sqrt{\log\sqrt{\frac{1}{\epsilon}}},
\end{equation}
where $M=\big(Y_c - \langle {Y} \rangle - \alpha \big)_+$ measures the design margin and $U=D$ provides an unambiguous definition and measure of uncertainty. The ratio $\text{CF}$ of margin to uncertainty measures the confidence that can be placed on the design as is referred to as confidence factor. The design criterion eq.~(\ref{eq:certcrit}) simply requires that the confidence in the design, as measured by the confidence factor, be greater than a minimum value.

\subsection{Optimal uncertainty quantification (OUQ)}
The McDiarmid's approach for UQ in eqs.~(\ref{eq:bound}) and (\ref{yuS4aF}) is attractive because it requires limited but tractable information on input variables (i.e., independence and intervals), response functions (i.e., sub-diameters) and performance measures (i.e., mean response). A question of theoretical and practical importance concerns whether it is possible to obtain an optimal bounds on the probability of failure using the same given information. Other related questions concerns the possibility of using other information than sub-diameters and mean output.  These questions have been addressed by Owhadi et al.~\cite{Owhadi:2013}, and we proceed to summarize their main results for completeness. Assume that we want to certify the safety of a system and the criterion is given by 
\begin{equation} 
    \mathbb{P}[F(X) \geq Y_c]
    \leq
    \epsilon
\end{equation}
based on the information that $X \equiv (X_1,\dots,X_N)$, $X_1,\dots,X_N$ are independent, $X \in I$ and that $\sup | F(\hat{x}_i, x_i) - F(\hat{x}_i, x_i')| \le D_i$, $\mathbb{E}[F] \le 0$. As a result, the optimal bound $\mathcal{U}(\mathcal{A}_{MD})$ on the probability of failure $\mathbb{P}[F(X) \geq Y_c]$ is the solution of the following optimization problem
\begin{equation} 
    \mathcal{U}(\mathcal{A}_{MD}) = 
    \sup_{(G,\mu) \in \mathcal{A}_{MD}} 
    \mu [G(X) \geq Y_c]
    \label{eq:OUQ}
\end{equation}
where
\begin{equation} 
	\mathcal{A}_{MD} = \left\{ (G,\mu) \middle\vert
	\begin{array}{c}
	 G: I_1 \times \cdots \times I_N \rightarrow \mathbb{R} \\
	 \mu \in  \mathcal{M}(I_1) \otimes \cdots \otimes \mathcal{M}(I_m) \\
	 \mathbb{E}_\mu [G] \le 0 \\
	 \sup |G(\hat{x}_i, x_i) - G(\hat{x}_i, x_i')| \le D_i
	\end{array}
	\right\}
	\label{eq:inf}
\end{equation}
and $\mathcal{M}(I_i)$ denotes the set of probability measures on $I_i$.

In practical applications, the available information does not determine a unique solution of $(G,\mathbb{P})$ but instead provides an information set $\mathcal{A}$. This set applies constraints on $(G,\mathbb{P})$ and hence consists of all possible values of $(G,\mathbb{P})$. As a result, the optimal uncertainty quantification (OUQ) (Owhadi et al.~\cite{Owhadi:2013}) aims to find the optimal bounds on probabilities given such set of information about the uncertainties. These bounds are calculated as extreme values of well-deﬁned optimization problems corresponding to extremizing probabilities of deviation subject to the constraints imposed by the known information. As a result, McDiarmid's concentration-of-measure approach in eqs.~(\ref{eq:bound}) and (\ref{yuS4aF}) provides an upper bound on $\mathcal{U}(\mathcal{A}_{MD})$. 

Although the optimization problem eq.~(\ref{eq:OUQ}) needs to be solved in infinite-dimensional spaces of measures and functions and therefore is extremely large, under general moment and independence conditions, Owhadi et al.~\cite{Owhadi:2013} have shown that they have finite-dimensional reductions. An application of OUQ that is relevant to the present work concerns the development of explicit and optimal concentration inequalities of the McDiarmid's type. Namely, considering the information given in eq.~(\ref{eq:inf}) and assuming $D_1 \ge \cdots \ge D_N$, if $Y_c \ge \sum_{i=1}^{N-2} D_i + D_N$, the optimal bound is given by
\begin{equation} 
    \mathcal{U}(\mathcal{A}_{MD}) = 
    \begin{cases}
    0 & \text{if} ~ \sum_{i=1}^ND_i \le Y_c, \\
    \frac{(\sum_{i=1}^N D_i - Y_c)^N}{N^N\prod_{i=1}^N D_i} & \text{if} ~ \sum_{i=1}^N D_i - ND_N \le Y_c \le \sum_{i=1}^N D_i, \\
    \frac{(\sum_{i=1}^k D_i - Y_c)^k}{k^k \prod_{i=1}^k D_i} & \text{if, for} ~ k \in \{1,...,N-1\},\\
    & \sum_{i=1}^k D_i - kD_k \le Y_c \le \sum_{j=1}^{k+1} D_i-(k+1)D_{k+1}. 
    \end{cases}
    \label{eq:OUQ_MD}
\end{equation}
We also refer to Owhadi et al.~\cite{Owhadi:2013} for detailed derivation of eq.~(\ref{eq:OUQ_MD}). The resultant optimal bounds provide a means of improving on the simple McDiarmid's bounds that are taken as the basis for the present work. Since the bounds are optimal, further improvements inevitably require information other than or in addition to system sub-diameters and mean performance.

\section{Numerical experiments}
\label{sec:sim}

\begin{figure}[!ht]
\centering
\includegraphics[width=4.0in]{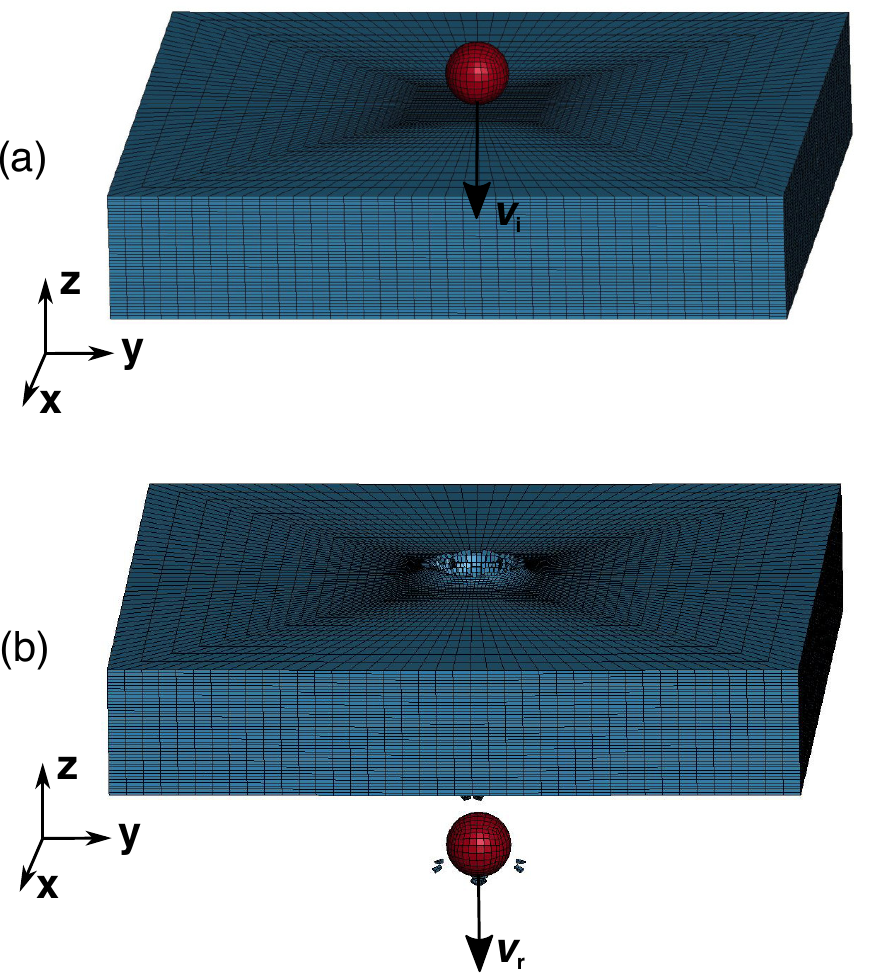}
\caption{Schematic of the ballistic problem. (a) Initial setup of the projectile/plate system. (b) The system after perforation. The impact and the residual velocities of the projectile are denoted by $v_{\mathrm{i}}$ and $v_{\mathrm{r}}$, respectively.}
\label{fig:setup}
\end{figure}

We now proceed to quantify the uncertainties of the constitutive and failure properties in the modeling of a ballistic impact problem, using the UQ strategies described in the foregoing. Specifically, the target is an AZ31B Mg alloy plate and the projectile is a steel ball, as shown in Fig.~\ref{fig:setup}(a). The residual velocity of the projectile is considered as the objective of interest in the numerical experiments, which can be considered as a metric to measure the performance of the plate subject to high-speed impact.  Fig.~\ref{fig:setup}(b) visualizes the system after penetration. We assume that all uncertainties arise from an imperfect characterization of the mechanical response of the plate. As a simple scenario, we further assume that, under the conditions of interest, the plate is well-characterized by the Johnson-Cook plasticity and fracture models~\cite{johnson1983constitutive, johnson1985fracture}, but the parameters of the two models are uncertain. Specifically, they must be allowed to vary over certain ranges in order to cover the experimental data with prescribed probability. For simplicity, the projectile is assumed to be rigid and uncertainty-free. 

\subsection{Material modeling}

We assume that the constitutive behavior of the plate is characterized by an appropriately calibrated Johnson-Cook plasticity model~\cite{johnson1983constitutive},
\begin{equation}
    \sigma \big( \epsilon_p, \dot{\epsilon}_p, T \big)
    =
    \big[A + B \epsilon_p^n \big]
    \big[1 + C \ln \dot{\epsilon}_p^* \big]
    \big[1 - {T ^*} ^m \big],
\label{eq:johnsoncook}
\end{equation}
where $\sigma$ is the true Mises stress, $\epsilon_p$ is the equivalent plastic strain, $\dot{\epsilon}_p$ is the plastic strain rate, and $T$ is the temperature. The normalized plastic strain rate $\dot{\epsilon}_p^*$ is defined as
\begin{equation}
    \dot{\epsilon}_p^* := \frac{\dot{\epsilon}_p}{\dot{\epsilon}_{p0}},
\label{eq:estar}
\end{equation}
where $\dot{\epsilon}_{p0}$ is a reference strain rate. The model also uses the normalized temperature
\begin{equation}
    T^* := \frac{T - T_0}{T_m - T_0},
\label{eq:tstar}
\end{equation}
where $T_0$ is a reference temperature and $T_m$ is the melting temperature. The model parameters are: $A$, the yield stress; $B$, the strain-hardening modulus; $n$, the strain-hardening exponent; $C$, the strengthening coefficient of strain rate; and $m$, the thermal-softening exponent.

We further assume that the failure behavior of the plate is well-described by the Johnson-Cook fracture model~\cite{johnson1985fracture}. Specifically, the damage of an element is defined on a cumulative damage parameter
\begin{equation}
    E
    =
    \sum\frac{\Delta \epsilon}{ \epsilon^f},
\label{eq:damageparam}
\end{equation}
where the summation is conducted over time steps and $\Delta \epsilon$ is the plastic strain increment in each time step. $ \epsilon^f$ denotes the strain at fracture which is given by 
\begin{equation}
    \epsilon^f
    =
    \big[E_1 + E_2 \exp(E_3 \sigma^*)\big]
    \big[1 + E_4 \ln \dot{\epsilon_p}^* \big]
    \big[1 + E_5 T^* \big],
\label{eq:fracturestrain}
\end{equation}
where $E_1$, $E_2$, $E_3$, $E_4$ and $E_5$ are material damage constants. ${\epsilon_p}^*$ and $T^*$ are defined in eqs.~\ref{eq:estar} and \ref{eq:tstar}, respectively. $\sigma^*$ is the ratio of the pressure $p$ divided by the von-Mises equivalent stress, i.e.,
\begin{equation}
    \sigma^*
    =
    \frac{p}{\sqrt{\frac{1}{2}\big[ (\sigma_1-\sigma_2)^2 + (\sigma_2-\sigma_3)^2 
    + (\sigma_3-\sigma_1)^2 \big]}},
\end{equation}
where $\sigma_1$, $\sigma_2$ and $\sigma_3$ are the principle stresses. Based on this damage model, fracture takes place when the damage parameter $E$ reaches the value of $1$.

We regard the set $X \equiv (A, B, n, C, m, E_1, E_2, E_3, E_4, E_5)$ of Johnson-Cook plasticity and damage parameters as the main source of uncertainty in the analysis. The estimated values and the bounds are tabulated in Table~\ref{tab:randparam}. In practice, the material parameters are derived from specific data sources, e.g., experiments or sub-grid simulations. We assume that these data are sufficient to determine confidence intervals for each parameter. In our calculations, we specifically use the AZ31B Mg alloy characterization of Hasenpouth \cite{hasenpouth2010tensile}, which, conveniently, includes the lower and upper bound of the $95$\% confidence intervals for Johnson-Cook plasticity parameters, i.e., $A$, $B$, $n$, $C$ and $m$.  In addition, regarding the Johnson-Cook damage parameters, the estimated values are provided by Feng {\it et al.}~\cite{feng2014constitutive}, but their bounds of confidence intervals are scarce in the literature. To this end, we add the similar level of uncertainty, i.e., $10.0$\%, to the estimated values to generate the lower and upper bounds, and the resultant values are also shown in Table~\ref{tab:randparam}.

\begin{table}[!ht]
\centering
\caption{Estimated values and bounds of AZ31B Johnson-Cook plasticity and damage parameters.}
\begin{tabular}{l l l l l}
\hline
\hline
Parameter & ~Estimated value & ~Lower bound  &  ~Upper bound & ~Uncertainty \\
\hline
$A$ (MPa)  &  $225.171$   &   $200.372$ &    $249.970$  & $+/-11.01$\% \\
$B$ (MPa)  &  $168.346$  &  $150.682$  &     $186.010$ & $+/-10.49$\% \\
$n$             &   $0.242$      &  $0.160$      &      $0.324$    & $+/-33.88$\% \\
$C$            &    	$0.013$     &   $0.012$     &       $0.014$    & $+/-7.69$\% \\
$m$           &    $1.550$      &   $1.523$     &       $1.577$     & $+/-1.74$\% \\
$E_1$         & $-0.35$         & $-0.385$    & $-0.315$         & $+/-10.0$\% \\
$E_2$        & $0.6025$      & $0.5423$    & $0.6628$       & $+/-10.0$\% \\ 
$E_3$        & $-0.4537$    & $-0.4991$   & $-0.4083$     & $+/-10.0$\% \\ 
$E_4$        & $0.4738$     & $0.4264$    & $0.5212$        & $+/-10.0$\% \\ 
$E_5$        & $7.2$            & $6.48$         & $7.92$            &   $+/-10.0$\% \\  
\hline
\end{tabular}
\label{tab:randparam}
\end{table}


\begin{table}[!ht]
\centering
\caption{Fixed system parameters of the plate used in the ballistic problem.}
\begin{tabular}{l l l l}
\hline
\hline
Plate (AZ31B Mg)                &     Value               & Unit     & Source               \\
\hline
Mass density                    &     $1.77$              & g/$\text{cm}^3$ & -       \\
Young's modulus                 &     $45.0$              & GPa & -                    \\
Poisson's ratio                 &     $0.35$              & -   & -                       \\
Specific heat                  & $1.75$               &  J/(K$\cdot$g)  & \cite{lee2013thermal} \\
Taylor-Quinney factor  & $0.6$ & - & \cite{kingstedt2019conversion}                \\
Spall strength & $1.5$ & GPa & \cite{farbaniec2016microstructural} \\
Gruneisen intercept & $4520.0$ & m/s  & \cite{feng2017numerical} \\
Gruneisen gamma & $1.54$ & -  & \cite{feng2017numerical} \\
Gruneisen slope $S_1$ & $1.242$ & -  & \cite{feng2017numerical} \\
Reference strain rate           &     $0.001$             & $\text{s}^{-1}$ & \cite{hasenpouth2010tensile}       \\
Reference temperature           &     $298.0$             & K & \cite{hasenpouth2010tensile}                      \\
Reference melt. temp.           &     $905.0$             & K & \cite{hasenpouth2010tensile}                      \\
Plate length/width                     &     $5.08$              & cm   & -                    \\
Plate thickness                    &     $0.953$              & cm          & -            \\
\hline
\end{tabular}
\label{tab:paramplate}
\end{table}

\begin{table}[!ht]
\centering
\caption{Fixed system parameters of the projectile used in the ballistic problem.}
\begin{tabular}{l l l}
\hline
\hline
Projectile (Steel)           &     Value                &  Unit             \\
\hline
Mass density                   &     $7.83$              & g/$\text{cm}^3$       \\
Young's modulus                &     $210.0$               & GPa                   \\
Poisson's ratio                &     $0.30$               &  -                      \\
Diameter                       &     $0.476$               & cm                    \\
\hline
\end{tabular}
\label{tab:paramprojectile}
\end{table}

\subsection{Forward solver}

For a given realization of the system parameters, the ballstic imapct problem is solved using the explicit dynamics solver available within the commercial finite element analysis software package LS-DYNA \cite{hallquist2007ls}. The initial conditions of the computational model are shown in Fig.~\ref{fig:setup}(a). The plate is resolved using $185,600$ elements, while the number of elements for the projectile is $3,584$. All the elements are linear hex, single point integration with careful hourglass control. Elements are refined in the impact region of the plate. This number and distribution of elements are enough to make the calculations converged. No constraint is applied on the plate. All simulations were ran for $30~\mu\text{s}$ before termination. This simulation duration is sufficiently long to allow for either the penetration of the projectile through the plate, or the rebound and separation of the projectile from the plate. The time-step size is adaptive and determined by the critical size of elements. Element erosion is used to characterize material failure. Additionally, the calculations are adiabatic with the initial temperature set at room temperature. The equation-of-state, which controls the volumetric response of the material, is assumed to be of the Gruneisen type. The values of the fixed parameters of the plate and the projectile used in the calculations are tabulated in Tables~\ref{tab:paramplate} and \ref{tab:paramprojectile}, respectively.

\subsection{Ballistic behavior}

\begin{figure}[!ht]
\centering
\includegraphics[trim=0.7in 2.5in 0.7in 2.5in,clip,width=0.6\textwidth]{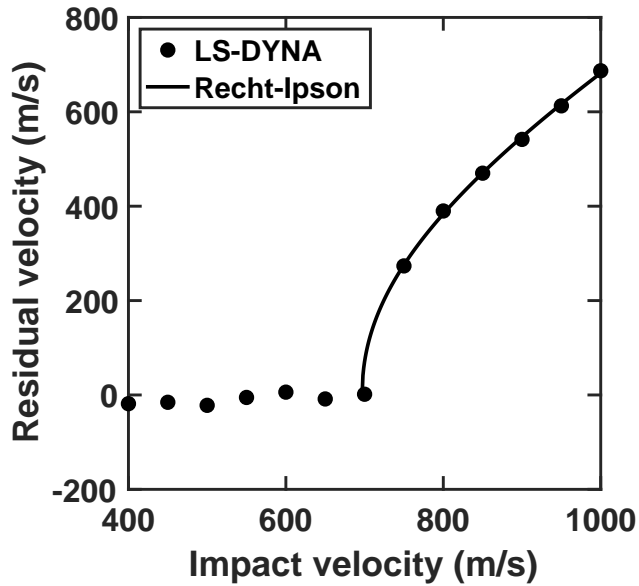}
\caption{Residual velocity as a function of impact velocity.}
\label{fig:resivel}
\end{figure}

We begin by evaluating the ballistic property of the plate subject to normal impact, using one representative group of the Johnson-Cook plasticity and damage parameters listed in Table~\ref{tab:randparam}. Fig.~\ref{fig:resivel} shows the residual velocity of the projectile, as a function of impact velocity ranging from $400$~m/s to $1000$~m/s. In order to estimate the ballistic limit, we employ the Recht-Ipson model~\cite{recht1963ballistic}, which provides an analytical expression based on the conversation of energy and momentum. Specifically, the model has the form of
\begin{equation}
    v_\text{r}
    =
    a \big( v_\text{i}^p - v_\text{bl}^p \big) ^{1/p},
\label{eq:Recht-Ipson}
\end{equation}
where $a$, $p$ and $v_\text{bl}$ are fitting parameters. Specifically, $v_\text{bl}$ represents the ballistic limit of the plate. The Recht-Ipson model is fitted to the residual perforation velocities calculated by LS-DYNA, and the values of the fitting parameters are tabulated in Table~\ref{tab:Recht-IpsonParam}. For comparison, the resulting Recht-Ipson curve is also plotted in Fig.~\ref{fig:resivel}. It is notable that the LS-DYNA results conform to the form of the Recht-Ipson model. The coefficient of determination $R^2$ is extremely close to $1$, also showing that the Recht-Ipson predictions well fit the LS-DYNA data. The predicted value of the ballistic limit by LS-DYNA calculations is $696.9$~m/s.  Moreover, although the parameter $p$ is fitted to the numerical data, its value is very close to that given in the original Recht-Ipson model (i.e., $p=2$). 

\begin{table}[!ht]	
\begin{threeparttable}[b]
\centering
\caption{Values of Recht-Ipson parameters for the AZ31B Mg plate under consideration.}
\begin{tabular}{l l l l l}
\hline
\hline
Parameter  & $v_\text{bl}$ (m/s) & $a$       & $p$         & $R^2$ \\
\hline
Value         & $696.9$                 & $0.916$ & $2.120$ & $0.999$ \\
\hline
\end{tabular}
\label{tab:Recht-IpsonParam}
\end{threeparttable}
\end{table}

\begin{figure}[!ht]
\centering
\includegraphics[width=6.5in]{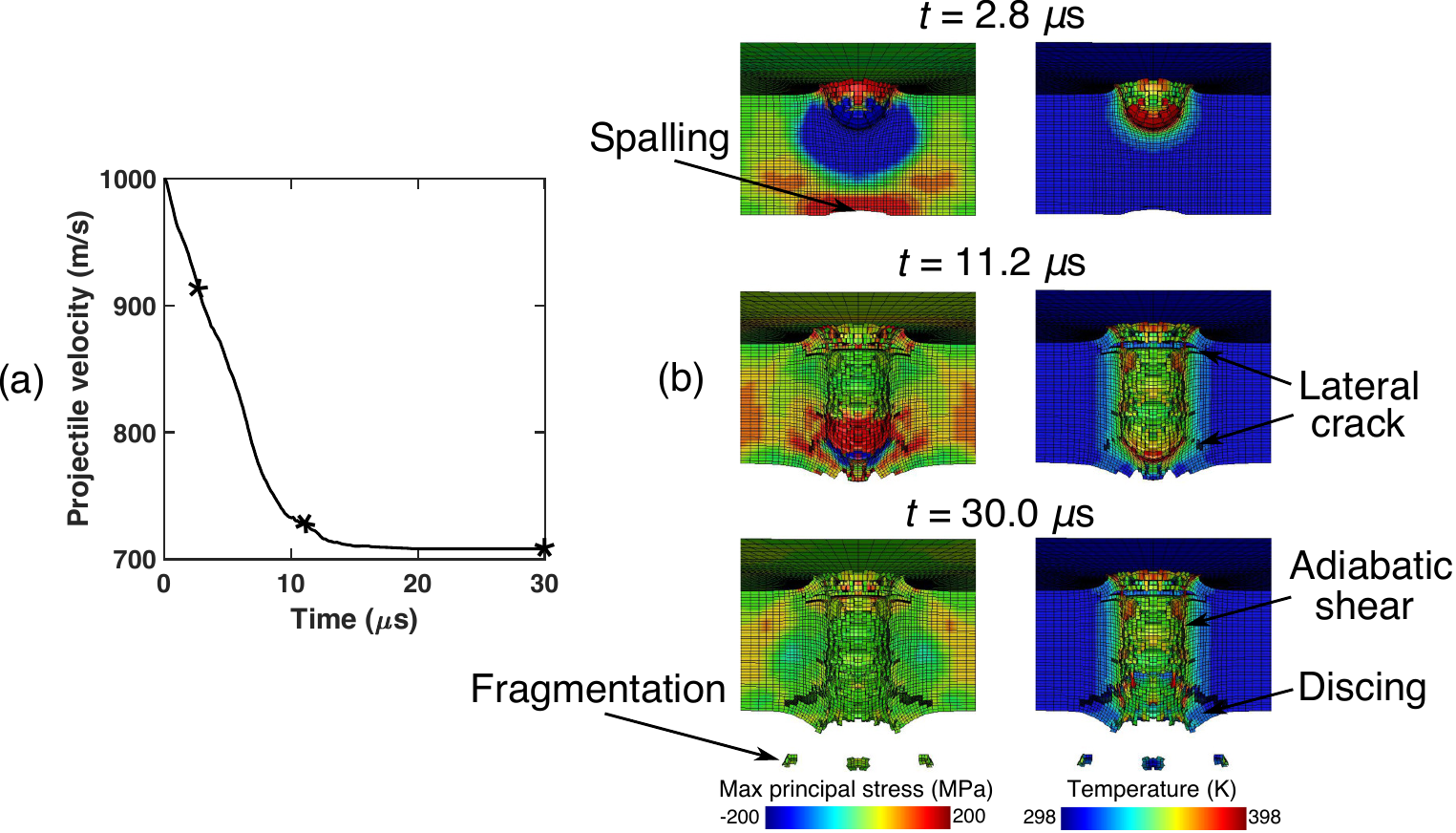}
\caption{Normal impact of the AZ31B Mg plate with the initial velocity of $1000$~m/s. (a) Time history of the projectile velocity. (b) Middle cross-section of the impact region. In Subfigure (a), the asterisks highlight the three time instances shown in Subfigure (b). In Subfigure (b), the projectile is removed for the sake of clarity.}
\label{fig:failure}
\end{figure}

We proceed to examine material failure mechanisms involved in the LS-DYNA calculations. Fig.~\ref{fig:failure} shows the time history of the projectile velocity and the snapshots of the impact regions at three time instances of the calculation concerned with the normal strike with the impact velocity of $1000$~m/s. Specifically, Fig.~\ref{fig:failure}(b) shows level contours of maximum principal stress and temperature at the same moments in time. As may be seen from Fig.~\ref{fig:failure}(a), the velocity of the projectile decreases smoothly during the penetration process. The penetration completes at around $20~\mu\text{s}$, and afterwards the velocity equals to a constant, i.e., the residual velocity. In addition, from Fig.~\ref{fig:failure}(b) it is noteworthy that the LS-DYNA solver, when equipped with the element erosion criterion, is capable of capturing several well-known failure modes for materials under high-speed impact.  Overall, the plate is perforated mainly by a conventional plugging mechanism. The impact energy is released by the large plastic work around the shear rupture that separates the cavity from the rest of the plate. The temperate field also peaks at the cavity boundary, resulting in thermal softening of the plate. This thermal softening in turn facilitates and promotes localization of deformation, eventually resulting in plugging formation. In addition to the shear plugging, spalling occurs in the region of high maximum principal stress near the backface of the plate at $2.8~\mu\text{s}$, due to the interaction between two reflecting tensile waves. Then at $11.2~\mu\text{s}$, because of the bending and stretching of the plate, several lateral cracks have appeared, which are parallel to the plane of the plate. As a result of the lateral cracks, discing happens near the backface of the plate at $30.0~\mu\text{s}$. The calculation also shows fragmented materials, which is typical for brittle materials such as AZ31B Mg alloys.

\subsection{UQ analysis}

Computing sub-diameters requires a constrained optimization over the space of input variables in order to determine the largest deviation in the performance measure. To this end, we employ a genetic algorithm (GA), which, as a global and derivative-free optimization method, provides the greatest flexibility in applications to non-linear problems. Another advantage of the GA is its high degree of concurrency. In particular, each iteration of the solution algorithm can be evaluated independently across multiple processors. In calculations, we employ the DAKOTA Version $6.7$ software package~\cite{adams2014dakota} of the Sandia National Laboratories. We choose throughout a fixed population size of $64$. One seed in the initial population is generated by setting the two repeated optimization variables associated with the sub-diameter at the two limits of that parameter range, with the remaining optimization variables set at the mid-span of their respective ranges. The remaining individuals in the initial population are selected randomly. We find that this initial setup accelerates the convergence of the GA iterations. Additionally, the crossover and the mutation rates are $0.8$ and $0.25$, respectively. The numbers pf parents and offspring of each generation are $64$ and $48$, respectively. More details about the computational framework can be found at Ref.~\cite{sun2020rigorous}.

\begin{figure}[!ht]
\centering
\includegraphics[width=6.5in]{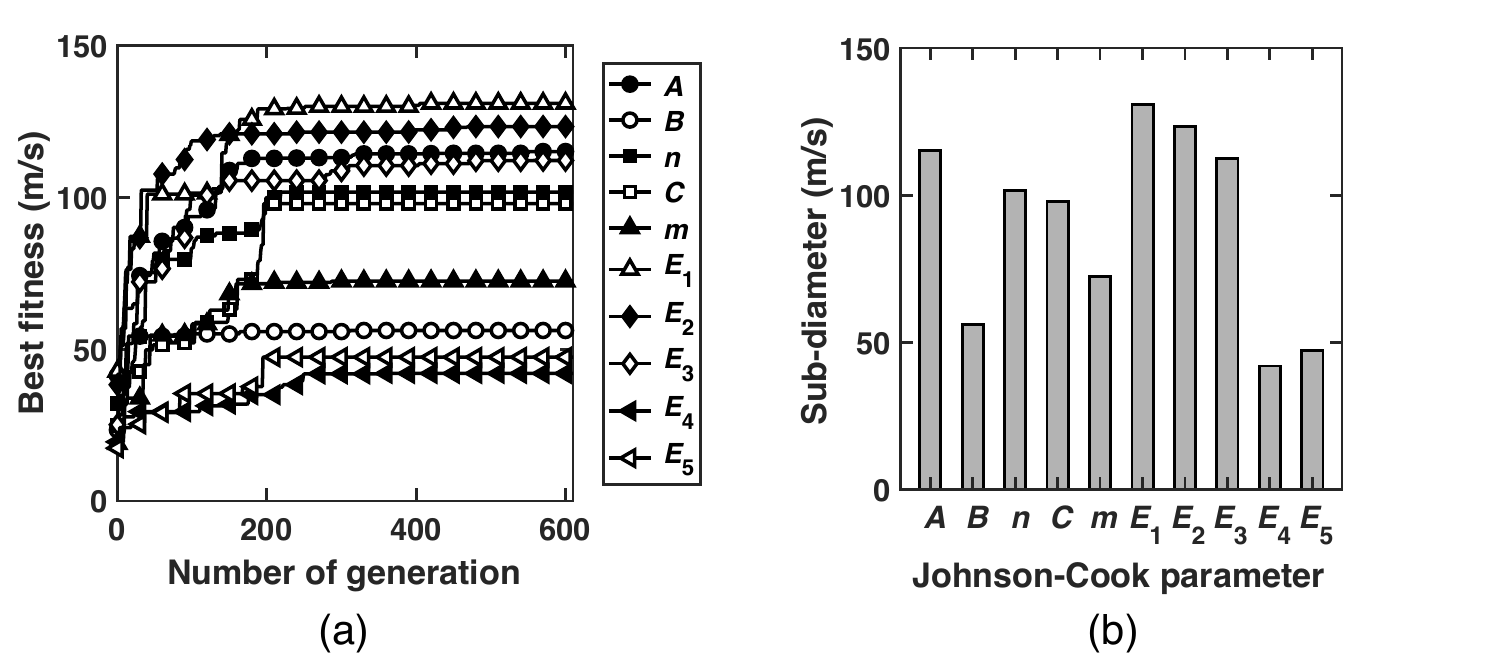}
\caption{Uncertainty quantification for the ballistic problem. (a) History of the best fitness in GA calculations. (b) Sub-diameters of random parameters in Johnson-Cook plasticity and fracture models.}
\label{fig:SubDiam}
\end{figure}

\begin{table}[!ht]	
\centering
\caption{Sub-diameters of Johnson-Cook parameters. The total diameter is $300.97~\text{m/s}$, to which the plasticity and fracture models contribute $203.91$~m/s and $221.37$~m/s, respectively.}
\begin{tabular}{l l}
\hline
\hline
Johnson-Cook parameter  &   Sub-diameter (m/s)  \\
\hline
$A$  & $115.07$  \\        
$B$ & $56.15$   \\  
$n$ & $101.73$  \\  
$C$ & $97.94$  \\  
$m$ & $72.41$ \\   
$E_1$ & $130.95$ \\     
$E_2$ & $123.30$ \\ 
$E_3$ & $112.46$ \\ 
$E_4$ & $42.00$ \\ 
$E_5$ & $82.48$ \\   
\hline
\end{tabular}
\label{tab:SubDiam}
\end{table}

As aforementioned, the Johnson-Cook parameters $A$, $B$, $n$, $C$, $m$, $E_1$, $E_2$, $E_3$, $E_4$ and $E_5$ are assumed to be uncertain and known to be within intervals only. Specifically, in the UQ analysis we investigate the worst-case scenario for all operating conditions under consideration, i.e., the largest velocity of $1000$~m/s with normal impact. We use the residual velocity of the projectile as the performance measure of the system, which provides an excellent measurement for the behavior of the AZ31B Mg plate under high-speed impact. Fig.~\ref{fig:SubDiam}(a) shows the best fitness of each generation in the GA calculations. It is notable that, in spite of high non-linearity and irregularity of the fracture model and the contact condition used in the present ballistic problem, all GA calculations start to converge to a maxima after $200$ to $300$ generations. For the sake of comparison, Fig.~\ref{fig:SubDiam}(b) compares the sub-diameters computed for each of the material parameters of the Johnson-Cook plasticity and fracture models. The numerical values of the sub-diameters are also tabulated in Table~\ref{tab:SubDiam}. It is noteworthy that the sub-diameters are measured in the unit of the performance measure. Therefore, they all have the same unit. A direct consequence of this property is that the sub-diameters can be compared and rank-ordered, which in turn provides a quantitative metric of the relative contributions of the parameters to the overall uncertainty of the response. From Fig.~\ref{fig:SubDiam}(b), we thus deduce this rank-ordering to be $E_1>E_2>A>E_3>n>C>m>B>E_5>E_4$, with the parameters $E_1$, $E_2$ and $A$ that contribute the most to the uncertainty, $B$, $E_5$ and $E_4$ the least and $E_3$, $n$, $C$ and $m$ intermediate. It is also notable that the contribution rank-ordering is different from the order of percentile variation of the random variables. For instance, compared to other parameters, the parameter $n$ has the greatest percentile variation, cf.~Table~\ref{tab:randparam}, but contributes only modestly to the total uncertainty in the system performance. This example evinces how relative uncertainties cannot be directly deduced from the variability of the input parameter in general, but also depend critically on the non-linear sensitivity of the system response to the parameters.  

Another finding from the sub-diameters is the model-dependent analysis of uncertainty contributions. As mentioned before, we assume that the material uncertainty comes from two separate models, i.e., Johnson-Cook plasticity and fracture models. Based on the sub-diameters listed in Table~\ref{tab:SubDiam}, the corresponding total diameter is $300.97~\text{m/s}$. The uncertainty contribution by the plasticity model, which includes the parameters $A$, $B$, $n$, $C$ and $m$, is $203.91$~m/s. By contrast, the fracture model, characterized by $E_1$, $E_2$, $E_3$, $E_4$ and $E_5$, contributes $221.37$~m/s to the total uncertainty. Therefore, for the selected random parameters and their ranges under consideration, the Johnson-Cook fracture model contributes more to the uncertainty of the ballistic behavior of the AZ31B Mg alloy plate.

\begin{figure}[!ht]
\centering
\includegraphics[width=6.5in]{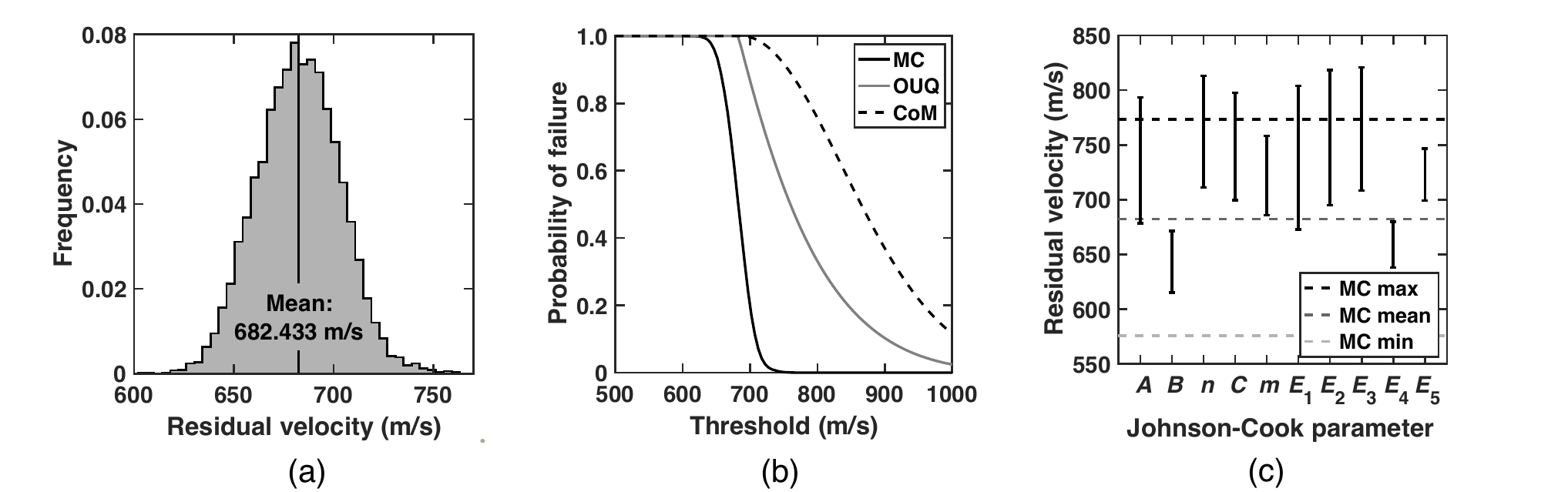}
\caption{Comparison with MC sampling. (a) Histogram of output distribution by MC sampling. (b) Probability of failure computed from direct MC sampling and bounded by OUQ and CoM. (c) Residual velocities at the bounds of sub-diameters.  In subfigure (c), the mean, maximum and minimum values of sampling results by MC are also shown.}
\label{fig:MC}
\end{figure}

As already mentioned, the present approach is predicated on probability inequalities as a means of bounding uncertainties. Evidently, the tighter the bound the better the design. However, increasing tightness comes at increasing computational expense, which sets forth a trade-off between economy of design and computability. Simple probability inequalities, such as McDiarmid's~\cite{mcdiarmid1989method}, supply a working compromise between tightness and computational complexity. However, it is both interesting and useful to investigate the tightness of the bounds and the attendant conservativeness of the designs. To this end, Fig.~\ref{fig:MC} shows comparisons between Monte Carlo (MC) sampling, concentration-of-measure (CoM) inequality and optimal uncertainty quantification (OUQ) for the projectile/plate system with impact velocity $v_\text{i}=1000~\text{m/s}$ and normal attack. The MC sampling is performed over the ranges of the random parameters, using $1.2\times10^4$ samples with Latin hypercube scheme and uniform distribution. Specifically, Fig.~\ref{fig:MC}(a) shows the sampled distribution of the residual velocity $Y$. We compute $\langle Y \rangle = 682.433$~m/s. Using this mean value, the upper bounds of the probability of failure through CoM and OUQ are shown in Fig.~\ref{fig:MC}(b), compared with an estimate using the MC sampling. As expected, both the CoM and OUQ bounds lie uniformly above than the MC estimate, which illustrates the conservative character of the bounds and, by extension, of the corresponding designs. For the problem with the information of system sub-diameters and mean performance, OUQ provides the optimal bound. As a result, the OUQ bound is tighter than CoM bound, as also shown in Fig.~\ref{fig:MC}(b). Moreover, Fig.~\ref{fig:MC}(c) shows the two values of the residual velocity at the bounds of the sub-diameter for each random variable. The mean, maximum and minimum values of the sampled values of the residual velocity are also shown in Fig.~\ref{fig:MC}(c). Notably, the residual velocities associated with the sub-diameters of $A$, $n$, $C$, $E_1$, $E_2$ and $E_3$ are beyond the scope of MC data. These random parameters also contribute significantly to the total uncertainty in the performance of the plate. Thus, due to the sharp and irregular features involved in the contact and fracture problem, the random sampling method is not able to capture those extreme cases. By contrast, our UQ framework is capable of executing tests {\it on demand} over the entire operating space of random parameters, and hence provides required data for more rigorous uncertainty quantification and safer design.

\begin{figure}[!ht]
\centering
\includegraphics[width=6.5in]{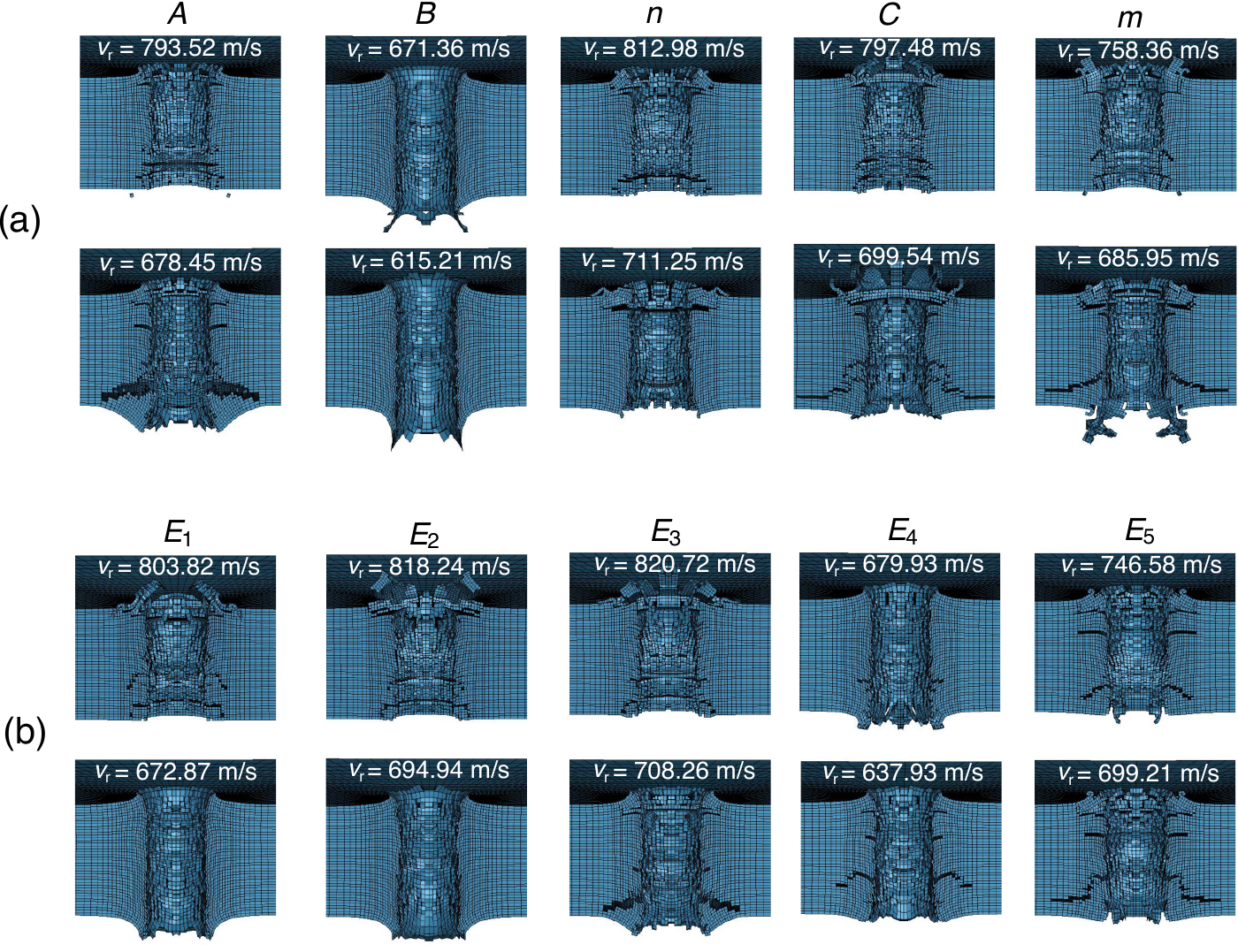}
\caption{Snapshots of cavity after perforation at the bounds of sub-diameter for each random parameter. (a) Plasticity parameters. (b) Damage parameters.}
\label{fig:penetr}
\end{figure}

The uncertainties considered in the present work come from the plasticity and fracture models, which characterize the strength and toughness of the materials, respectively. In order to explore the material regime in which the random parameters cause the largest deviation of the residual velocity, we extract the snapshots of the perforation cavity at the two bounds of the sub-diameter for each random variable, as shown in Fig.~\ref{fig:penetr}. Specifically, Fig.~\ref{fig:penetr}(a) and (b) visualize the profile of the cavity for the plasticity and fracture parameters, respectively. As may be seen from Fig.~\ref{fig:penetr}(a), the plasticity parameters result in the largest output uncertainty in different material regimes. Specifically, the behavior of the plate is more sensitive to the parameters $A$, $n$, $C$ and $m$ when the plate is relatively brittle, whereas the parameter $B$ has a significant effect in the regime of high toughness. Regarding the fracture parameters in Fig.~\ref{fig:penetr}(b), it is notable that the parameters $E_1$, $E_2$ and $E_4$ have the largest uncertainty contribution in the range crossing both the ductile and brittle regimes. By contrast, $E_3$ and $E_5$ have a more significant contribution to the total uncertainty in the brittle regime.

\section{Summary and concluding remarks}
\label{sec:concl}

We have presented the implementation of a UQ framework to assess the effects of constitutive and fracture properties on the performance of Mg alloys subject to high speed impact. The quantification towards uncertainties is achieved by determining the largest deviation in the performance measure resulting from a finite variation in the corresponding input random variables. Both strategies of McDiarmid's CoM inequality and OUQ have been harnessed to calculate conservative upper bounds on the probability of failure, and then the safety of the system can be certified and designed rigorously using the QMU strategy. The McDiarmid's CoM inequality has a relatively simple formulation and hence can be easily employed, whereas the optimal UQ is able to provide the optimal bounds by leveraging all known information of uncertainties and measurements in spite of its complex equations. The uncertainties of the constitutive and fracture properties in the present work arise from the partial information of model parameters, determined either by experiments or by experience of experts.

Several significant findings afforded by the calculations are noteworthy. For our specific ballistic impact of AZ31B and the given intervals of uncertainties, the sub-diameters of the Johnson-Cook constitutive and fracture parameters are ordered as $E_1>E_2>A>E_3>n>C>m>B>E_5>E_4$. Obviously, this ordering emphasizes the specific parameters where improvements are best targeted. Such a clear set of relationships also characterizes the constitutive and failure regime of interest, especially when compared against other boundary and/or initial conditions. For example, regarding the fracture models, the parameters related to the quasistatic behaviors, i.e., $E_1$, $E_2$ and $E_3$, contribute more significantly to the ballistic performance of AZ31B compared to the strain-rate-related parameter $E_4$ and temperate-related  parameter $E_5$. Therefore, the corresponding microscopic properties determining $E_1$, $E_2$ and $E_3$ need to be adjusted in order to greatly improve the ballistic performance of the material. On the other hand, if a new fracture model needs to be developed for AZ31B, the part that affects its quasistatic behavior should gain more attention. Moreover, the material failure mechanisms for AZ31B involved in the impact tests bear emphasis, including spalling, plugging, discing and fragmentation, and our ballistic tests also conform to the form of the Recht-Ipson model.

We close with a comparison between the adopted UQ approach and Bayesian strategies. Bayesian inference, based on the Bayes’ theorem, has been introduced as one of the main tools for UQ of the computational models, mainly due to the relative simplicity of implementation and the rigor of the resulting Bayesian analysis. Nevertheless, the quantification of uncertainties is conducted via calculating high-dimensional integrals that are very intractable or even impossible to evaluate analytically through conventional integration techniques, let alone those significantly complicated response functions that are achieved by existing open-source codes or commercial software. One numerical way to solve those integrals is Monte-Carlo sampling, which can become impractical if the probability of failure is small, i.e., if failure is a rare event, and if one-call of forward calculation is costly, as shown in the numerical tests in this paper. By way of contrast, in the adopted method the response function can effectively be regarded as a black-box and the effort required for the computation of the uncertainty bounds only depends on the solution of sub-diameters and therefore is independent of the size of the probability of failure. In addition, the most commonly used priors in engineering problems are uniform and normal distributions. However, the strong influence of priors on the outcome of the inference process is also one of the most significant criticisms of Bayesian frameworks. By contrast, the CoM and OUQ approaches only requires the intervals of uncertain parameters and then provides rigorous bounds on the output uncertainties that bracket all the possible results led by all the probability measures in such intervals. If more information of uncertainties and system, such as moments of random variables, is given, the OUQ approach is able to obtain tighter bounds by leveraging such information.

\section*{Acknowledgments}

XS gratefully acknowledges the support of the University of Kentucky through the faculty startup fund.

\bibliography{mybibfile}
\bibliographystyle{unsrt}

\end{document}